\documentclass[english,aps]{revtex4}
\usepackage[T1]{fontenc}
\usepackage{amsmath}
\usepackage{graphicx}
\usepackage{amssymb}

\makeatletter



\usepackage{babel}
\makeatother
\begin{document}

\title{Negative Entropy Production in Oscillatory Processes }

\author{Stephen R. Williams }

\email{swilliams@rsc.anu.edu.au}

\author{Denis J. Evans}

\email{evans@rsc.anu.edu.au}

\author{Emil Mittag}

\email{mittag@rsc.anu.edu.au}

\affiliation{Research School of Chemistry, The Australian National University,
Canberra, ACT 0200, Australia}

\begin{abstract}
Linear irreversible thermodynamics asserts that the instantaneous
local spontaneous entropy production is always nonnegative. However
for a viscoelastic fluid this is not always the case. Given the fundamental
status of the Second Law, this presents a problem. We provide a new
derivation of the Second Law, from first principles, which is valid
for the appropriately time averaged entropy production allowing the
instantaneous entropy production to be negative for short intervals
of time. We show that time averages (rather than instantaneous values)
of the entropy production are nonnegative. We illustrate this using
molecular dynamics simulations of oscillatory shear. 
\end{abstract}
\maketitle
Linear irreversible thermodynamics asserts that in local equilibrium
\cite{key-16}, the spontaneous entropy production per unit time,
per unit volume, the so-called entropy source strength $\sigma(\mathbf{\mathbf{r}},t)$,
cannot be negative \cite{key-1,key-3}. Further it states that the
entropy source strength is a sum of products of irreversible thermodynamic
fluxes $J_{i}$ and forces $X_{i}$,

\begin{equation}
\sigma(\mathbf{r},t)=\sum J_{i}(\mathbf{r},t)X_{i}(\mathbf{r},t)\geq0\label{entropy source strength}\end{equation}
where the source strength is calculated at a position $\mathbf{r}$
and a time $t$. For steady state processes close to equilibrium Eq.
\ref{entropy source strength} is clearly correct. However, for processes
that involve time dependent or oscillatory thermodynamic forces, in
viscoelastic materials close to equilibrium, Eq. \ref{entropy source strength}
is incorrect. These problems are usually resolved by separating the
fluxes into {}``storage and loss'' components \cite{key-4}. Such
a separation is not derived from first principles and is process specific.
A second problem with Eq. \ref{entropy source strength} is that it
is restricted to the near equilibrium, linear response, regime. Further
from equilibrium the definition of entropy and therefore temperature
remains an unsolved problem \cite{key-11}. Until very recently there
has been no known generalization of Eq. \ref{entropy source strength}
to the far from equilibrium regime.

Recent advances, which are of broad interest \cite{key-12,key-13},
have cast new light on these issues. In particular we will use the
Jarzynski Equality (JE) \cite{Jarzynski PRL,Jarzynski PRE} to derive
a variation on Eq. \ref{entropy source strength} from first principles.
In order to proceed we first define the conjugate flux $\mathbf{J}$,
for a system which is driven away from equilibrium by a dissipative
field $\mathbf{F}_{e}$, in terms of the rate that work is done on
the system \cite{evans-morriss},\begin{equation}
\mathbf{J}V\cdot\mathbf{F}_{e}\left(t\right)=-k_{B}T\Omega\left(t\right)\equiv k_{B}T\Lambda\left(\mathbf{\Gamma}\left(t\right)\right)-\frac{d}{dt}H_{0}\left(\mathbf{\Gamma}\left(t\right)\right).\label{JVFe}\end{equation}
Here $V$ is the system volume, $k_{B}$ is Boltzmann's constant,
$T$ is the temperature of the synthetic thermostat or a large heat
reservoir \cite{williams PRE 04}, $\Omega(t)$ is the dissipation
function as defined for the Evans-Searles Fluctuation Theorem \cite{evans-searles adv phys},
$\mathbf{\Gamma}\left(t\right)$ is the phase space vector, $H_{0}$
is the internal energy, $k_{B}T\Lambda$ gives the rate that heat
is lost from the system to the thermostat or reservoir \cite{williams PRE 04}
and $\Lambda$ is the phase space compression factor \cite{evans-morriss}.
Eq. \ref{JVFe} is defined for arbitrary field strengths and in the
linear response regime it gives the entropy production, \begin{equation}
-\langle\mathbf{J}V\cdot\mathbf{F}_{e}\left(t\right)\rangle/T=k_{B}\left\langle \Omega(t)\right\rangle =\int_{V}d\mathbf{r}\,\sigma\left(\mathbf{r},t\right)+O\left(F_{e}^{4}\right),\label{entropy production}\end{equation}
where $\left\langle ...\right\rangle $ denotes an ensemble average.

The JE has been shown \cite{Jarzynski PRL,Jarzynski PRE} and proved
\cite{Evans Mol Phys 03} for the time reversible thermostatted dynamics
we employ below. It has also been experimentally verified \cite{Bustamante sci,collin nat}.
The JE gives the change in Helmholtz free energy $\Delta A$ for a
system which has undergone a nonequilibrium process starting from
an initial ($t=0$) equilibrium distribution of phases $f_{1}(\mathbf{\Gamma},0)\propto exp[-\beta H(\mathbf{\Gamma},\lambda(0))]$
and initial Hamiltonian $H(\mathbf{\Gamma},\lambda(0))$ to a final
Hamiltonian $H(\mathbf{\Gamma},\lambda(\tau))$ at $t=\tau$ and then
relaxed to a new equilibrium. The dynamical proof \cite{Evans Mol Phys 03}
requires that the dynamical pathway $\mathbf{\Gamma}(t)$ be such
that the distribution of phases at $t=\tau$ can subsequently relax
to a final canonical distribution at $t=\infty$, $f_{\infty}(\mathbf{\Gamma},\infty)\propto exp[-\beta H(\mathbf{\Gamma},\lambda(\tau))]$.
The parametric transformation of the Hamiltonian is complete at a
finite time $t=\tau$, by which time the system is not expected to
have fully relaxed to the new equilibrium. JE states,

\begin{equation}
\left\langle \exp\left(-\beta W_{\tau}\right)\right\rangle =\exp\left(-\beta\Delta A\right).\label{eq:JE}\end{equation}
where $W_{\tau}=\int_{0}^{\tau}ds\dot{W}(s)$ is the Jarzynski work
function and $\dot{W}(t)\equiv[\frac{d}{dt}H_{0}(\Gamma(t),\lambda(t))]-k_{B}T\Lambda(\mathbf{\Gamma}(t))$,
\cite{Evans Mol Phys 03}. Although $H_{0}(\mathbf{\Gamma}(\tau),\lambda(\tau))\neq H_{0}(\mathbf{\Gamma}(\infty),\lambda(\tau))$
it is clear that, $W_{\tau}=W_{\infty}$.

The proof of the JE \cite{Evans Mol Phys 03} requires that the two
systems be connected by a path $1\rightarrow2$ and its inverse path
$2\rightarrow1$. The proof does not put restrictions on the time
dependence of the path. The parametric change in the Hamiltonian from
$\lambda(0)\rightarrow\lambda(\tau)$ may in addition contain work
due to the system being driven by a dissipative external field \cite{Evans Mol Phys 03}.
If the work is solely due to a dissipative external field ($\dot{\lambda}=0\;\;\forall\;\; t$)
then the rate of work $\dot{W}$ will be the same as that given by
Eq. \ref{JVFe} (ie $\dot{W}=k_{B}T\Omega$).

The JE allows a first principles proof of the Second Law Inequality
(SLI) from the equations of motion as first shown by Jarzynski \cite{Jarzynski PRE}.
By combining Eq. \ref{eq:JE} with the mathematical identity $\exp(x)\geqslant1+x$
we have, \begin{equation}
e^{-\beta\Delta A}=\langle e^{-\beta(W_{\tau}-\langle W_{\tau}\rangle)}\rangle e^{-\beta\langle W_{\tau}\rangle}\geq e^{-\beta\langle W_{\tau}\rangle}.\label{SLI proof}\end{equation}
Noting that $e^{x}$ is a monotonically increasing function we derive
the Clausius Inequality \begin{equation}
\left\langle W_{\tau}\right\rangle \geq\Delta A,\:\;\forall\;\:\tau\geq0.\label{Clausius}\end{equation}
If the system of interest is a fluid and if the transformation involves,
say, a shearing deformation or perhaps the translation of one particle
through a fixed distance, then clearly $\Delta A=0$, we may treat
the field as external $\dot{\lambda}=0$, and thus, by Eq. \ref{JVFe},
$W_{\tau}=k_{B}T\bar{\Omega}_{\tau}\tau=-\int_{0}^{\tau}ds\,\mathbf{J}\left(s\right)V\cdot\mathbf{F}_{e}\left(s\right)$
where $\bar{\Omega}_{\tau}\equiv\frac{1}{\tau}\int_{0}^{\tau}ds\,\Omega(s)$.
The SLI then follows,

\begin{equation}
\left\langle \bar{\Omega}_{\tau}\right\rangle =-\frac{1}{\tau k_{B}T}\int_{0}^{\tau}ds\,\left\langle \mathbf{J}\left(s\right)V\cdot\mathbf{F}_{e}\left(s\right)\right\rangle \geq0,\:\;\forall\:\;\tau\geq0,\:\forall\:\;\mathbf{F}_{e}(s)\label{SLI}\end{equation}
which forms a generalization of Eq. \ref{entropy source strength}
that is valid at arbitrary field strengths and derived from the equations
of motion. A significant difference between Eq. \ref{SLI} \& Eq.
\ref{entropy source strength} is that Eq. \ref{entropy source strength}
applies to instantaneous values whereas Eq. \ref{SLI} applies to
time averages of the entropy production starting at $t=0$ from an
initial canonical distribution. We note that Eq. \ref{SLI} has previously
been derived under a more restrictive set of conditions from the Evans
Searles Fluctuation Theorem \cite{key-5}.

We decided to test this prediction using nonequilibrium molecular
dynamics simulations of shear flow in a fluid. We consider the case
of sinusoidal shear applied to a viscoelastic fluid. We employ the
Lees-Edwards (sliding brick) periodic boundary conditions along with
the so-called SLLOD equations of motion for planar Couette flow \cite{evans-morriss},\begin{align}
\dot{\mathbf{q}}_{i} & =\frac{\mathbf{p}_{i}}{m}+\mathbf{i}\dot{\gamma}(t)q_{yi}\label{EOM}\\
\dot{\mathbf{p}}_{i} & =\mathbf{F}_{i}-\mathbf{i}\dot{\gamma}(t)p_{yi}-\alpha\mathbf{p}_{i}\nonumber \end{align}
where $\dot{\gamma}\equiv\partial u_{x}/\partial y$ is the strain
rate, $\mathbf{p}_{i}$ is the peculiar momentum taken relative to
the streaming velocity $u_{x}(y)=\mathbf{i}\dot{\gamma}(t)y$ and
$\alpha(t)$ is a Gaussian thermostat multiplier which holds the kinetic
temperature fixed \cite{evans-morriss}. For Couette flow with a constant
strain rate the SLLOD equations of motion are known to give an exact
description of steady adiabatic planar Couette flow for arbitrary
values of the strain rate \cite{evans-morriss}. For time dependent
shear flows it is known that the SLLOD equations give an exact description
of such flows in the linear response regime for both adiabatic and
thermostatted flows \cite{evans-morriss}. For high shear, oscillatory
flows it is not known whether SLLOD is exact but it is widely assumed
this is so.

To apply the SLI, Eq. \ref{SLI}, to the equations of motion Eq. \ref{EOM},
we note that the equilibrium Hamiltonian or internal energy is $H_{0}=\sum p_{i}^{2}/2m+\Phi$
where $\Phi$ is the total interparticle pair potential. The dissipation
function, which in the linear response regime gives the entropy production
Eq. \ref{entropy production}, may then be obtained from,

\begin{equation}
k_{B}T\Omega\left(t\right)=-\dot{\gamma}VP_{xy}=-\dot{\gamma}\Bigl(\sum\limits _{i=1}^{N}\frac{p_{xi}p_{yi}}{m}-\sum\limits _{i<j}^{N}F_{xij}q_{yij}\Bigr),\label{entropy equation}\end{equation}
 where $N$ is the number of particles, $P_{xy}$ is the $xy$ element
of the pressure tensor, $F_{xij}$ is the $x$ component of the pairwise
additive force on particle $i$ due to particle $j$, $q_{yij}$ is
the $y$ component of the vector connecting their centers and $p_{xi}$
is the $x$ component of the momentum of particle $i$. So in this
case for $\mathbf{J}V\cdot\mathbf{F}_{e}$ we have $\mathbf{J}=P_{xy}$
and $\mathbf{F}_{e}=\dot{\gamma}$. 

Simulations of oscillatory shear were carried out on a fluid in three
dimensions using the pair potential $\phi_{ij}=\epsilon[(\sigma/r_{ij})^{12}-(\sigma/r_{ij})^{6}+0.25]\;\forall\; r_{ij}<2^{1/6}$,
with the volume and temperature held constant. The equations of motion
were solved using a fourth order Runge-Kutta algorithm with a time
step $\triangle t=0.001$. The number density is $\rho=N\sigma^{3}/V=0.95$,
$N=108$, $k_{B}T=\epsilon$ and the time unit is $\sqrt{m\sigma^{2}/\epsilon}$
throughout. For times $t\leq0$ the system was in equilibrium. The
initial equilibrium configurations were obtained by sampling an equilibrium
trajectory, at time intervals of 5, which was generated by solving
Eq. \ref{EOM} with $\dot{\gamma}=0$. Starting from these initial
configurations a total of $5\times10^{5}$ nonequilibrium oscillatory
Couette flow trajectories were computed: $\dot{\gamma}(t)=\dot{\gamma}_{0}\sin(\omega t),\quad t>0$
with $\omega=4\pi$ and $\dot{\gamma}_{0}=0.2$. The duration of the
nonequilibrium trajectories was 2. %
\begin{figure}
\resizebox{8.5cm}{!}{\includegraphics{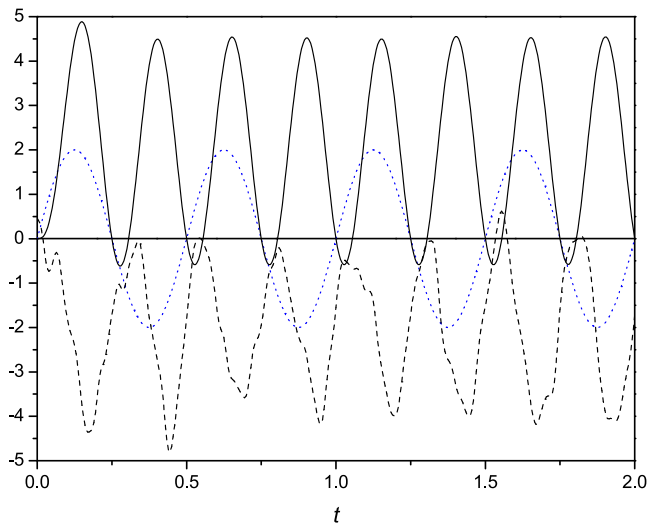}}

\caption{The response of the system to the oscillatory shear, in the linear
response regime, which began at time $t=0$ following Eq. \ref{EOM}.
The solid line is the entropy production, $\left\langle \Omega(t)\right\rangle =-\dot{\gamma}P_{xy}(t)V/\epsilon$.
The dashed line is the instantaneous rate of heat absorbed by the
thermostat $\left\langle dQ/dt\right\rangle /\epsilon$. The dotted
line gives the strain rate $\dot{\gamma}(t)$ multiplied by a factor
of 10. The data for the heat exchange is considerably more noisy than
the other data. Clearly $\left\langle \Omega(t)\right\rangle $ is
at times negative.}
\end{figure}
\begin{figure}
\resizebox{8.5cm}{!}{\includegraphics{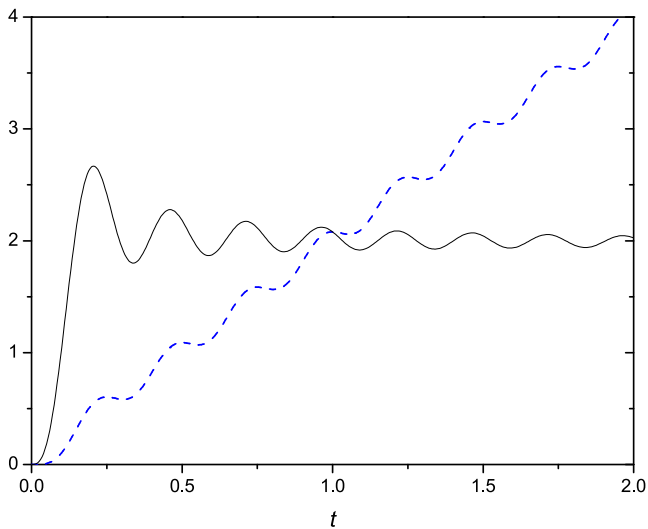}}

\caption{The dashed line is the integral, $\left\langle \bar{\Omega}_{t}\right\rangle t$,
which is nonmonotonically increasing but always nonnegative. The solid
line (black) is the time average $\left\langle \bar{\Omega}_{t}\right\rangle $
which is also always nonnegative. }
\end{figure}

In Fig. 1 the dissipation function, $\left\langle \Omega(t)\right\rangle $
in units of $k_{B}$ (which was calculated by ensemble averaging all
of the nonequilibrium trajectories) is plotted as a function of time
along with the ensemble averaged rate of heat absorbed by the thermostat,
$\dot{Q}(t)=-3Nk_{B}T\left\langle \alpha(t)\right\rangle =k_{B}T\left\langle \Lambda\right\rangle $
in units of $\epsilon$. Also shown is the strain rate, $10\times\dot{\gamma}(t)$.
The initial transients in the response decay very rapidly. It is clear
that the response of the fluid is viscoelastic: there is a phase lag
between $P_{xy}(t)$ and $\dot{\gamma}(t)$ due to the relatively
high frequency $\omega$. This shows that the dissipation function
(or at weak fields, equivalently, the entropy production Eq. \ref{entropy production})
is negative within certain intervals of time even though the system's
response is linear. This effect is not due to the amplitude of the
shear rate $\dot{\gamma}_{0}$ being large. This is clearly at odds
with the traditional view from irreversible thermodynamics. In contrast
Eq. \ref{SLI} is satisfied at all times as may be seen in Fig. 2
where the integral $\left\langle \bar{\Omega}_{t}\right\rangle t$
and the time average $\left\langle \bar{\Omega}_{t}\right\rangle $
are plotted as a function of time $t$. 

In summary we have shown the assertion of linear irreversible thermodynamics
that the instantaneous entropy production is always nonnegative is
incorrect for the case of time dependent viscoelastic fluids even
if they are in the linear response regime close to equilibrium. The
Second Law Inequality Eq. \ref{SLI} derived from the Jarzynski Equality
states that the time integral (starting from $t=0$) of the ensemble
averaged dissipation function cannot be negative for arbitrary integration
times and arbitrary field strengths (of course in the weak field limit
the dissipation function is equal to the entropy production Eq. \ref{entropy production}).
This inequality requires that the Helmholtz free energy of the corresponding
equilibrium system does not change. For planar shear this is a necessary
condition for the fluid state, which by definition cannot support
a constant stress. For a solid, undergoing oscillatory nonplastic
deformation, the equilibrium free energy would depend on the deformation
and the Clausius Inequality Eq. \ref{Clausius} will need to be used
rather than the Second Law Inequality Eq. \ref{SLI}. 

Lastly we note that the Second Law Inequality is a \textit{macroscopic}
consequence of the Jarzynski Equality and of the Evans Searles Fluctuation
Theorem \cite{key-5}. All previously derived consequences of the
JE and the Fluctuation Theorem were microscopic in nature. The Second
Law Inequality in the form Eq. \ref{SLI}, has important consequences
in applications such as atmospheric physics where the principle of
maximum entropy in nonequilibrium states has been employed \cite{key-15}.

We thank the Australian Research Council for financial support, the
Australian Partnership for Advanced Computing, and Debra J. Searles,
Edie Sevick and Chris Jarzynski for helpful discussions. DJE thanks
Siegfried Hess for reminding him of this problem.


\begin{thebibliography}{10}
\bibitem{key-16}Local equilibrium requires that the local thermodynamic
potentials are the same function of thermodynamic state variables
that they are in total equilibrium \cite{key-3}: in the case presented
here the same function of the thermostat temperature and the number
density. For an isotropic fluid variables such as the pressure, the
internal energy and the entropy do not change to linear order in the
amplitude of the external field regardless of the time dependence.
Thus the local equilibrium requirement and the linear response regime
are formely equivalent. This can be shown from response theory \cite{evans-morriss}.
To linear order the average of the $xy$ element of the pressure tensor
$P_{xy}$, for a process which begins at $t=0$, is given by the Green
Kubo relation\[
\langle P_{xy}(t)\rangle=-\beta V\int_{0}^{t}ds\dot{\gamma}(t-s)\langle P_{xy}(0)P_{xy}(s)\rangle_{0}+O(\dot{\gamma}_{0}^{3})\]
where the notation $\left\langle \ldots\right\rangle _{0}$ denotes
that the correlation function is determined for a system in equilibrium.
Thermodynamic potentials are scalar variables which, in an isotropic
fluid, do not change to linear order. If we denote a scalar variable
as $B(t)$ then, in an isotropic fluid, its cross correlation function
with a tensor element has the the property, $\left\langle B(0)P_{xy}(t)\right\rangle _{0}=0\;\;\forall\:\: t$,
due to symmetry. Thus to linear order in the external field scalar
variables do not change, \begin{align*}
\langle B(t)\rangle & =-\beta V\int_{0}^{t}ds\,\dot{\gamma}(t-s)\langle B(0)P_{xy}(s)\rangle_{0}+\langle B(0)\rangle+O(\dot{\gamma}_{0}^{2})\\
 & =\langle B(0)\rangle+O(\dot{\gamma}_{0}^{2}).\end{align*}
Equivalently one may expand the distribution function $f_{F_{e}}(\boldsymbol{\Gamma},t)$
as a Taylor series in the external field,\[
f_{F_{e}}(\boldsymbol{\Gamma},t)=f_{0}(\boldsymbol{\Gamma},t)+f_{1}(\boldsymbol{\Gamma},t)F_{e}+O(F_{e}^{2}),\]
where $f_{0}(\boldsymbol{\Gamma},t)$ is the equilibrium distribution
function, and arrive at the same conclusion. This is shown in detail
using Enskog theory in Ch IX, \S 6 of ref. \cite{key-3}.

\bibitem{key-1}D. Kondepudi and I. Prigogine, \textit{Modern Thermodynamics}
(Wiley, New York, 1998); See especially Eq. (15.2.3). 

\bibitem{key-3}S. R. De Groot and P. Mazur, \textit{Non-equilibrium
Thermodynamics} (Dover, New York, 1984). 

\bibitem{key-4}P. J. Daivis and M. L. Matin, J. Chem. Phys. 118 (2003)
11111. 

\bibitem{key-11}J. R. Dorfman, \textit{An Introduction to Chaos in
Nonequilibrium Statistical Mechanics} (Cambridge University Press,
Cambridge, 1999). 

\bibitem{key-12}C. Bustamante, J. Liphardt and F. Ritort, Phys. Today
\textbf{58}(7), 43 (2005).

\bibitem{key-13}D. Ruelle, Phys. Today 57(5) (2004) 48.

\bibitem{Jarzynski PRL}C. Jarzynski, Phys. Rev. Lett. 78 (1997) 2690. 

\bibitem{Jarzynski PRE}C. Jarzynski, Phys. Rev. E 56 (1997) 5018.

\bibitem{evans-morriss}D. J. Evans and G. P. Morriss, \emph{Statistical
Mechanics of Nonequilibrium Liquids} (Academic, London, 1990). 

\bibitem{williams PRE 04}S. R. Williams, D. J. Searles and D. J.
Evans, Phys. Rev. E 70 (2004) 066113.

\bibitem{evans-searles adv phys}D. J. Evans and D. J. Searles, Adv.
Phys. 51 (2002) 1529.

\bibitem{Evans Mol Phys 03}D. J. Evans, Mol. Phys. 101 (2003) 1551.

\bibitem{Bustamante sci}J. Liphardt et. al., Science, 296 (2002)
1832.

\bibitem{collin nat}D. Collin et. al., Nature, 437 (2005) 231.

\bibitem{key-5}D. J. Searles and D. J. Evans, Aust. J. Chem. 57 (2004)
1119. 

\bibitem{key-15}R. Lorenz, Science, 299 (2003) 837.
\end{thebibliography}
\end{document}